\documentstyle[epsfig,natbib2,natbibmnfix]{mn}
\newcommand{\be}{\begin{equation}}
\newcommand{\ee}{\end{equation}}
\newcommand{\bea}{\begin{eqnarray}}
\newcommand{\eea}{\end{eqnarray}}
\newcommand{\bc}{\begin{center}}
\newcommand{\ec}{\end{center}}
\def\spose#1{\hbox to 0pt{#1\hss}}
\newcommand{\lta}{\mathrel{\spose{\lower 3pt\hbox{$\mathchar"218$}}
     \raise 2.0pt\hbox{$\mathchar"13C$}}}
\newcommand{\gta}{\mathrel{\spose{\lower 3pt\hbox{$\mathchar"218$}}
     \raise 2.0pt\hbox{$\mathchar"13E$}}}

\newcommand{\dgr}{^\circ}

\def\H0{$H_0 = 100~h~$km\,s$^{-1}$\,Mpc$^{-1}$}
\def\msun{$h^{-1}{\rm M}_{\odot}$}
\newcommand{\hmpc}{$h^{-1}$Mpc}
\newcommand{\mhmpc}{\,h^{-1}{\mbox{Mpc}}}

\newcommand{\ea}{et~al.}

\bibpunct[,]{(}{)}{;}{a}{}{,}

\title[Voids in LCRS]{Voids in the LCRS versus CDM Models}

\author[V.~M\"uller \ea]{V.~M\"uller$^1$\thanks{Email: vmueller@aip.de}, 
S.~Arbabi-Bidgoli$^1$\thanks{Email: sarbabi@aip.de},
J.~Einasto$^2$\thanks{Email: einasto@aai.ee},  
D.~Tucker$^3$\thanks{Email: dtucker@fnal.gov}
\\
$^1$Astrophysikalisches Institut Potsdam, An der Sternwarte 16, 
14482 Potsdam, Germany
\\
$^2$Tartu Observatory, 61602 T\~oravere, Estonia
\\
$^3$Fermilab, MS 127, Box 500, Batavia, IL 60510, USA
}

\begin{document}

\maketitle 

\begin{abstract} 

We have analyzed the distribution of void sizes in the two-dimensional
slices of the Las Campanas Redshift Survey (LCRS).  Fourteen
volume-limited subsamples were extracted from the six slices to
cover a large part of the survey and to test the robustness of the
results against cosmic variance.  Thirteen samples were randomly
culled to produce homogeneously selected samples.  We then studied the
relationship between the cumulative area covered by voids and the void
size as a property of the void hierarchy.  We find that the
distribution of void sizes scales with the mean galaxy separation,
$\lambda$. In particular, we find that the size of voids covering half
of the area is given by $D_{med} \approx \lambda + (12\pm3) \mhmpc$.
Next, by employing an environmental density threshold criterion to
identify mock galaxies, we were able to extend this analysis to mock
samples from dynamical $n$-body simulations of Cold Dark Matter (CDM)
models.  To reproduce the observed void statistics, overdensity
thresholds of $\delta_{th} \approx 0 \dots 1$ are necessary.  We have
compared standard (SCDM), open (OCDM), vacuum energy dominated
($\Lambda$CDM), and broken scale invariant CDM models (BCDM): we 
find that both the void coverage distribution and the two-point
correlation function provide important and complementary information
on the large-scale matter distribution. The dependence of the void 
statistics on the threshold criterion for the mock galaxy 
indentification shows that the galaxy biasing is more crucial for the 
void size distribution than are differences between the cosmological 
models.

\end{abstract}

\begin{keywords} 
cosmology:  dark matter -- galaxies: formation -- large scale structure 
of the universe.
\end{keywords}

\section{Introduction}

Voids in the distribution of galaxies were first noticed in early
studies of the large scale distribution of galaxies, \citet{JET78},
\citet{GT78}, \citet{TF78}, \citet{CR79}, and \citet{Ta79}.  The void
behind the Perseus-Pisces supercluster (\citet{JET78}) and the
Bo\"otes void (\citet{KOSS81}) both have diameters of about 70 \hmpc\/
(\H0).  The presence of voids was explained by \citet{EJS80} and
\citet{ZES82} by gravitational instability.  Matter disperses by
outflow from low-density regions (voids), and the rest remains there
in primordial form; in high-density regions, however, the matter
collapses and forms superclusters, filaments of galaxies, and
clusters.  The resulting picture resembles a `cellular' structure and
is well described by the pancake theory of \citet{Z70}.  The
evacuation of matter in voids has been traced back to the underlying
large-scale potential distribution by \citet{CMS93}, \citet{LS98}, and 
\citet{MDGM98}.  The galaxy distribution has also been characterized 
either as a `foam' of bubbles (\citet{dLGH86}) or as a `sponge-like 
network' of interlocking filaments and tunnels connecting overdense 
and empty regions (\citet{GMD86}).  More recent studies have shown 
that voids may be populated and subdivided into smaller voids by 
fainter galaxies, cp.\ \citet{Li95} and \citet{PHE97}.  Observationally, 
there is no doubt that large underdense regions which contain almost 
no galaxies are a common feature of the large scale structure. 
Furthermore, numerical simulations demonstrate that filaments and voids 
form in a hierarchy of scales for any physically reasonable model of 
structure formation (\citet{Me83}).

In the past, properties of voids have been characterized by the void
probability function (\citet{W79}) and by the statistics of void
diameters (\citet{EEG89}).  The void probability function has a clear
statistical interpretation; however, it falls off quickly below 20
\hmpc\/ as seen in the CfA catalog by \citet{VGH91}, in the 1.2 Jy IRAS
survey by \citet{Bou93}, and in the SSRS by \citet{dC94}.  Therefore,
it is not very sensitive to the matter distribution on large scales.
On the other hand, the distribution of diameters of maximum voids
describes better the distribution of galaxies and clusters on large
scales.  Numerical simulations within CDM models (\citet{EEGS91},
\citet{LW94}, \citet{JMBF94}, \citet{VGPH94}, \citet{Gh94}, and
\citet{Gh96}) have shown that CDM models can explain the observed void
probability function for a suitable cosmological model and a
corresponding bias model.  Appropriate models can also reproduce the
dependence of the void sizes on the mean galaxy density in the
catalogues.

So far, void statistics have been investigated using galaxy surveys
which cover a large area on the sky.  Up till now, the deepest surveys
employed in such analysis have had radial extents of about 150 \hmpc.
For this reason, the measured size of voids was mostly restricted by
the size of the survey volume.  That is, until now, there has never
been a statistically complete measurement of the void distribution;
previous measurements have all been truncated at the high end by the
survey depths.  The aim of this paper is to study void properties in
the Las Campanas Redshift Survey (LCRS), which has a depth of about
600 \hmpc\/ ($z \lta 0.2$).  This depth is large enough to contain a
sufficient number of voids for a statistical analysis; thus this
sample is better suited for the investigation of void properties over
a broader scale interval.  The price to be paid is that the survey
consists of 6 narrow strips on the sky, i.e.  it is effectively
2-dimensional.  We take this into account and project the galaxies on
the six central planes, and we perform a void analysis in two
dimensions.  To this aim, we apply the void finder algorithm developed
by \citet{KF91} and \citet{KM92} (similar algorithms can be found in
\citet{EP97}, \citet{EPdC97}, and \citet{EP00}).  The void finder puts
arbitrarily formed but approximately convex voids of maximal size into
the galaxy distribution.  Then we study the mean fractional area
covered by voids (\citet{KM92}), as it is a stable characteristic of
the void distribution.  We also compare the properties of voids in the
LCRS with voids in large simulations of CDM models with different
cosmological parameters.

The paper is organized as follows.  In Section 2 we describe the selection
of volume limited samples from the different slices of the LCRS and
analyze the distribution of void sizes in the selected data sets.  In
Section 3 we present results of numerical simulations of four cosmological
models, discuss the prescription for establishing mock catalogues, and
apply the void finder to them.  In section 4 we compare data and
simulations and draw our conclusions.

\section{Voids in volume limited subsamples of the LCRS}

The LCRS is the deepest redshift survey presently available
(\citet{LCRS96}).  The survey contains 24,518 galaxy redshifts in 3
slices in the northern and in 3 slices in the southern galactic
hemisphere.  Each slice extends $\approx 80 \dgr$ in right ascension
and $\approx 1.5 \dgr$ in declination.  The northern galactic
hemisphere slices are centered at $\delta = -3 \dgr, -6 \dgr, -12
\dgr$, and the southern at $\delta = -39 \dgr, -42 \dgr, -45 \dgr$.
Here, we select volume limited subsamples in the different slices,
correct for the non-uniform sampling rate, and project the galaxies to
the central plane.  Taking into account the dominant two-dimensional
geometry of the LCRS, we look for two-dimensional voids in these
planes.  Whether these voids are representative for voids in the
three-dimensional galaxy distribution will be discussed in a future
paper.

\begin{figure}
\label{fig1}
\psfig{figure=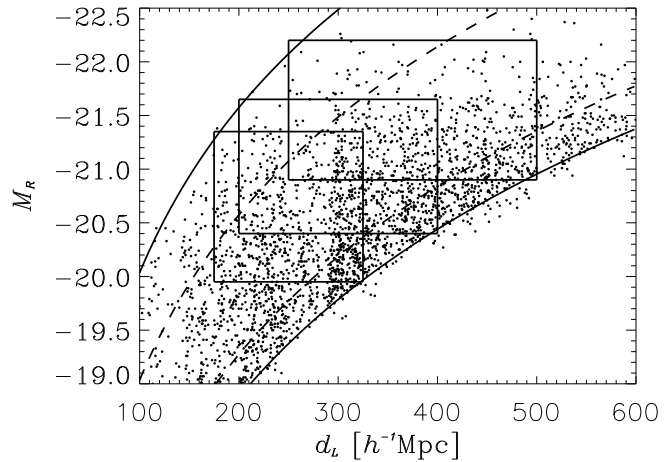,width=9cm}
\caption{Selection of the volume limited subsamples 7, 8, and 9 in the 
$\delta = -12 \dgr$ slice. The magnitude limits of $m_R = 15$ and 17.7 and 
of $m_R = 16$ and 17.3 for the high and low sampling rate are shown as 
solid and dashed curves, respectively.}
\end{figure}

\subsection{Selection of volume limited subsets}

The LCRS contains galaxies within apparent magnitude ranges of $16 <
m_R < 17.3$ and of $15 < m_R < 17.7$ for the 50 and 112-fiber fields,
respectively.  Therefore we must impose both lower and upper limits in
depth and absolute magnitude to define volume limited samples.  We
select lower and upper limits of the luminosity distances, $d_1$ and
$d_2$, respectively, which cover the well sampled region of the survey
(see, for instance, Fig. 1).  For determining luminosity distances
$d$, we employ the relation of \citet{M58} with $q_0=0.2$ and a
$k$-correction $k(z) = 2.5 \log(1+z)$, which is representative for the
galaxy mix in the LCRS (\citet{Li96}).  From the distance limits and
the apparent magnitude range of the survey, we get lower and upper
limits of the absolute magnitude $M > M_1$ and $M < M_2$, within which
galaxies are observed in the chosen distance range (Fig. 1).  In all
slices we select at least 2 absolute magnitude limited samples in
different magnitude ranges in order to test our results for a possible
dependence on absolute magnitude.  The corresponding limits for 14
data sets are listed in Table ~\ref{data}.

Each slice in the LCRS consists of a collection of $1.5\deg^2$ fields
with different sampling characteristics which depend both on the
instrument used to obtain the spectra (either a 50-fiber or a
112-fiber spectrograph) and on the local galaxy surface density within
that field.  Therefore, there are field-to-field sampling variations
within each slice.  To obtain homogeneously sampled slices, we
randomly dilute the higher sampled fields to the minimum sampling rate
in the corresponding slice, as is shown in the third column of Table
\ref{data}.  Only in the final set 14 of Table \ref{data} do we keep
all the galaxies.  We take this set as control sample to have both a
high galaxy number and a high surface density.  This set is used to
test the effects of random sampling.  The resulting galaxy numbers are
given in the 8th column; these vary strongly due to the different
magnitude ranges and sampling fractions.  Column 9 shows the mean
galaxy separation $\lambda = \sigma^{-1/2}$, where $\sigma$ is the
surface density of galaxies.  There are large differences in the
galaxy density and in the volume covered by the different data sets.
In particular, sets 7 and 8, which stem from the one slice which was 
observed entirely with the 112-fiber spectrograph, have a high
galaxy density and are expected to give statistically reliable
results.  We employ the dependence of the void statistics on the
galaxy density in our further analysis.  It should be remarked that we
keep the galaxy slices in a wedge-like geometry.  A similar geometry is
taken for the study of the mock samples, and we discuss the effect of
the geometry and of boundary effects below.

\begin{table*}
\begin{minipage}{120mm}
\caption{\label{data} Properties of the volume limited subsamples of LCRS}
\bc
\begin{tabular}{cccccccccccc}
set & slice &$f_{min}$& $M_1$ & $M_2$ & $d_1$  & $d_2$ & $N_{gal}$ &
$\lambda$ & $D_{med}$ & $D_{max}$ \\
    &       &     &         &         & \hmpc  & \hmpc &         &
\hmpc    &      \hmpc   &   \hmpc \\
 1  & -45   & .21 & -20.20  & -21.10  &   220  &  320  &   138   &
12.56    &     31.2    &  38.4   \\
 2  & -45   & .21 & -21.00  & -21.80  &   320  &  440  &   136   &
15.83    &     29.6    &  53.6   \\
 3  & -42   & .28 & -20.30  & -20.80  &   210  &  330  &   182   &
14.26    &     27.2    &  40.8   \\
 4  & -42   & .28 & -20.90  & -21.60  &   300  &  420  &   165   &
16.52    &     32.8    &  47.2   \\
 5  & -39   & .30 & -20.50  & -21.00  &   235  &  360  &   221   &
14.08    &     28.8    &  55.2   \\
 6  & -39   & .30 & -20.90  & -21.40  &   280  &  420  &   182   &
17.43    &     29.6    &  55.2   \\
 7  & -12   & .45 & -19.95  & -21.35  &   175  &  325  &   829   &
7.05     &     20.0    &  40.8   \\
 8  & -12   & .45 & -20.40  & -21.65  &   200  &  400  &   803   &
8.85     &     21.6    &  44.0   \\
 9  & -12   & .45 & -20.90  & -22.20  &   250  &  500  &   638   &
12.04    &     27.2    &  46.4   \\
10  &  -6   & .39 & -20.10  & -20.70  &   200  &  300  &   223   &
10.96    &     20.8    &  38.4   \\
11  &  -6   & .39 & -20.60  & -21.40  &   280  &  380  &   213   &
12.47    &     21.6    &  37.6   \\
12  &  -3   & .37 & -20.00  & -20.40  &   180  &  280  &   174   &
12.27    &     24.0    &  38.4   \\
13  &  -3   & .37 & -20.50  & -21.10  &   240  &  360  &   280   &
11.79    &     22.4    &  42.4   \\
14  & -12   &  -- & -20.40  & -21.65  &  200  &  400  &  1217  &
7.25    &     21.6    &  37.6   \\
\end{tabular}
\ec
\end{minipage}
\end{table*}

\subsection{Void statistics in the LCRS} 

\begin{figure}
\label{fig2}
\vspace{-0.2cm}
\psfig{figure=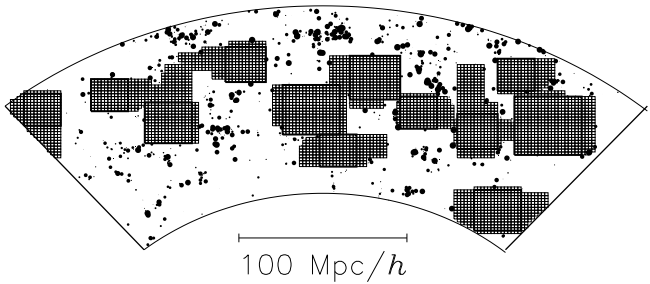,width=12.0cm}
\vspace{-1cm}
\caption{Large voids with extensions in set 7 selected from the
$\delta = -12 \dgr$ slice. Points represent the projected galaxy
positions and the point size is proportional to the galaxy's absolute
magnitude.}
\end{figure}

As illustrated in Fig. 2, we employ the algorithm of \citet{KF91} to
define voids in a two-dimensional plane of the LCRS.  To this aim we
construct a density field on a rectangular $1000 \times 1000$ grid
covering the survey plane, placed into a square of 800 \hmpc\/ on a
side.  The density field is defined by the galaxy number per cell, and
voids are connected samples of empty cells.  The algorithm places
maximal square boxes into the galaxy distribution.  As a next step,
extensions to each base void are constructed along all sides with the
restriction that each extension is a connected row of empty cells with
length exceeding two-thirds of the previous extension (or the base of
the square void).  This provides a good approximation of convex voids
as illustrated in Fig. 2.  In particular, it avoids the situation of
single voids consisting of different convex regions connected by
narrow tunnels.  After defining a void, the void cells are marked, and
voids with smaller base sizes are determined in the rest of the plane.
Unlike other algorithms, such as smoothing the density field and
fitting ellipsoids into underdense regions, we make no additional
assumptions about the void shape besides near convexity.  The void
finder which we have applied looks for completely empty regions in the
galaxy distribution.  Possibly this restriction may be circumvented in
a further development of the algorithm.

Of special importance for the void statistics is the treatment of the
boundary of the data sets.  After some trial and error, we decided to
count all cells outside of the survey region as occupied.  That means
that no voids are allowed to enter the boundary of the survey (see
Fig. 2).  This will restrict the size of some voids near the boundary,
and thus shift the void distribution slightly to smaller sizes $D$.
Tests with mock galaxy samples in larger volumes and in the survey
geometry show that this underestimation is less than 3\%.  We employ
similar geometries both in the observed data and in the mock samples
in order to be independent of this underestimation.  Nonetheless, the
effect should be taken into account in the quantitative evaluation of
the void sizes.

Here we look for the size distribution of voids, where the size is
measured by the length $D$ of the base voids.  We measure the
abundance of voids by the fraction of the slice area which is covered
by voids of a given size.  In Fig. 3 we show the cumulative
distribution of the coverage of the planes of the LCRS data sets.  The
smoothness of the histograms illustrates both the good statistics
which we get with our 14 data sets and the large depth of the LCRS.
The void distributions have obviously a similar shape.  The different
curves show that larger void sizes are typical for data sets with
larger mean galaxy separation $\lambda$ as given in Table \ref{data}.
A scaling of the void sizes with the galaxy density was already
noticed by \citet{RT84} who found that the maximum void size $D_{max}$
(with their void definition)
\be D_{max}\approx (2-3) n^{-1/3}. \ee 
For our two-dimensional data this corresponds to a scaling $\propto
\lambda = \sigma^{-1/2}$, which will be used in the following
analysis. Here we show the real physical sizes to demonstrate that
voids of (20 - 40) \hmpc\/ base sizes are typical for the bright
galaxies sampled in the LCRS.  Some voids have a size of up to 55
\hmpc.  They are rare, however, and their statistics are noisy.

\begin{figure}
\label{fig3}
\psfig{figure=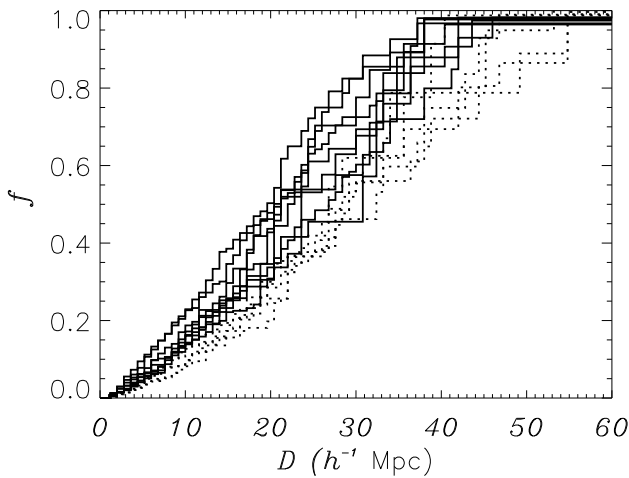,width=9cm}
\caption{Cumulative distribution of the fraction $f$ of the LCRS slices 
covered by voids of size $D$.  The solid lines show the better sampled 
sets 7, 14, 8, 10, 13, 12, 1, 9, 11 (from left to right) with 
$\lambda \lta 12$ \hmpc; the distributions for the other sets are dotted.}
\end{figure}

The basic factor that determines the void size distribution is the
mean galaxy surface density, $\sigma$, or, equivalently, the mean
galaxy separation $\lambda = \sigma^{-1/2}$.  We show in the last two
columns of Table \ref{data} the median and the maximum void size,
$D_{med}$ and $D_{max}$, respectively.  In Fig.  4, we show the
$\lambda$ dependence of the median and quartile values of the void
sizes -- i.e.  those values that lead to a 25\% , 50\%, and 75\%
coverage of the survey area if the void areas are added starting from
small sizes.  These values show the expected dependence on the galaxy
separation $\lambda$,
\be D \approx \nu\lambda + D_0, \ee 
with a residual void size $D_0$ for `zero mean separation' and a slope
$\nu$ for the size increase if a random subset of the data is studied.
The residual value $D_0$ is a somewhat formal quantity since the
minimum mean separation for bright galaxies as sampled in the LCRS
amounts a few Mpc.  Furthermore, $D_0$ is not well determined since it
lies well outside of the $\lambda$ range of the fit ($7 < \lambda <
17.5 \mhmpc$); note the quite large uncertainty of this value for the
data in Table \ref{fit}, in which the values for the typical void
sizes and the slopes $\nu$ are tabulated. The 1-$\sigma$ errors in the 
parameter ranges originate from the different data sets, i.e. they 
include statistical effects from different parts of the survey, from 
possible systematic effects in the void finder as using the wedge geometry, 
boundary effects, and also systematic effects due to different volumes 
and magnitudes in the different data sets. The last systematics is 
more intensely discussed below. It can be noted that a Poisson sample
leads typically to a slope $\nu \approx 2$ and to a residual value
$D_0$ consistent with zero, i.e. not unexpectedly, it is strongly different 
from the void statistic of the LCRS data.

A similar relation for the mean void size of a Poisson sample of
points in 3 dimensions, $D_{med} \approx 3\lambda$, was given in
\citet{Li95}.  The relation for the maximum void size in Eq.  (1) from
\citet{RT84} corresponds to $\nu \approx 2-3$ and no residual value,
which is typical for a Poisson point distribution.  Obviously, a quite
undersampled data set was used in their analysis.  Our relation in Eq.
(2) seems more typical for the void size distribution in a clustered
galaxy distribution, and an indication on a residual value is also
seen in the void distribution of \citet{Li95} (see Fig. 9 in their
paper, where the relation of the void size and the galaxy number is
shown).  The residual median void size, $D_0 \approx 12 \mhmpc$, is
taken as a typical size of voids in well sampled parts of the LCRS.
It substantially exceeds the corresponding values for the Poisson
samples given in the second row of Table \ref{fit}.  Also, the slope
of the fractional increase of the void size for diluted samples, $\nu
\approx 1$, is significantly different from that for the Poisson
samples, $\nu \approx 2$.  Obviously, the void sizes grow much more
quickly in diluted data sets for the random samples than for the
clustered data, and $\nu \approx 1$ is a typical result that must be
reproduced by simulated data sets.  It is also remarkable that the
inhomogeneous data set 14 lies on the same fit.  This means that some
inhomogeneity in the galaxy sampling does not destroy the typical
distribution of observed voids.

There are no major differences between the void distributions for
different bright galaxy samples, as the comparison of the void
statistics of data sets 1 and 12 with the much brighter data sets 9,
11, and 13 demonstrates (see Fig. 5 and Table 1).  Their median sizes
show no strong trend with the magnitude range, and only the maximum
void size of set 9 exceeds all the others due to its larger volume.
We do not regard the maximum void size as a reliable statistics for
our comparison of data and simulations.  In Fig. 5 we show the median
void sizes of the 14 data sets versus the lower limit of the
absolute magnitude range $M_1$.  There is a trend of increasing void sizes
for more luminous galaxies.  However, this trend is masked by
differences in the dilution factor, as seen by comparison of the data
sets sampled with high ($f_{min}>0.35$) and low ($f_{min}<0.35$)
sampling rate.  Thus we shall use in the following analysis the mean
galaxy separation in different data sets as the argument which
determines void sizes.  We studied Pearson's linear correlation $r$
(see, e.g. \citet{PTVF92}) of the median void size $D_{med}$ and the
mean galaxy separation $\lambda$, which is $r = 0.81$ with error
probability 0.04\%.  This indicates a clear correlation with high
reliability.  Taking the dependence of the median void size $D_{med}$
on the lower absolute magnitude limit $M_1$ and the sampling fraction
$f_{min}$, we get much weaker correlation coefficients of $r = 0.57$
and $r = 0.56$, with 3\% and 4\% error probability, respectively.  Hence,
basically, the first dependence, $D_{med}$ on $\lambda$, shows a tight
correlation.  This formal test shows that for the restricted range of
$M_1$ which could be tested with the given data the dependence of void
sizes on the absolute magnitude limit cannot be established reliably.

An other important effect which plays a role is the wedge-like
geometry of the LCRS data sets -- i.e, the LCRS slices are not
rectangular volumes, but wedges with an opening angle of $1.5\dgr$.
This leads to a slight density gradient over the survey plane and --
as shown by the scaling in Eq. (2) -- to somewhat smaller voids in the
more distant parts of the survey planes and to somewhat larger voids
in the less dense sampled parts of the nearby regions.  The addition
of small voids concerns less than half of the volume, whereas small
voids are less abundant in the other part due to the large volume
occupied by the big voids.  Tests with simulations show that the
cumulative void distribution in Fig. 3 is shifted to larger void sizes
$D$ by $(5 - 10)\%$.  This overestimate of void sizes also affects all
mean characteristics of the void distribution that are shown in Fig. 4
and given in Table \ref{fit}.  The same effect occurs in the mock
samples; i.e., this effect does not influence the comparison of the
data with models.  Tests have shown that it is more reliable to keep a
higher galaxy number in the volume limited samples rather than to
reduce further the galaxy number in order to get samples with constant
thickness.  Finally, we remarked above that the boundary effects of
the survey tend to slightly reduce the typical void size -- i.e., they
yield a competing effect to that of the slices' wedge-like geometry.
Even so, as mentioned above, this competing influence is smaller
(only $\approx 3 \%$).

\begin{table*}
\begin{minipage}{170mm}
\caption{\label{fit} Void distribution in data and mock samples for some 
 CDM models: model parameter and fits to the residual void size $D_0$ and 
 and slope $\nu$ of void size increase.}  

\begin{tabular}{cllccrrcccrc}

sets &$\Omega_m$& $h$  &$\sigma_{8}$&$\delta_{th}$& $\delta_{cr}$& 
 \multicolumn{2}{c}{median} & \multicolumn{2}{c}{lower quartile} & 
 \multicolumn{2}{c}{upper quartile}   \\

   & & & & &
   & $D_0$ & $\nu$ & $D_0$ & $\nu$ & $D_0$ & $\nu$ \\
   & & & & &
   & \hmpc &       & \hmpc &       & \hmpc &       \\

data     & & & & & 
         & $11.8 \pm 2.9$ & $1.1 \pm 0.2$ & $5.7 \pm 1.6$ & $0.9 \pm 0.1$ 
         & $16.8 \pm 2.9$ & $1.5 \pm 0.2$ \\

Poisson  & & & & &                     
         &  $0.9 \pm 1.0$ & $1.8 \pm 0.1$ & $0.9 \pm 0.3$ & $1.2 \pm 0.1$ 
         &  $2.1 \pm 0.5$ & $2.1 \pm 0.1$ \\ 
SCDM mock1 & 1 & 0.5 & 1.3 & -0.9 & 400
         &  $5.7 \pm 1.2$ & $1.6 \pm 0.1$ & $1.6 \pm 0.7$ & $1.0 \pm 0.1$  
         &  $6.8 \pm 1.2$ & $2.2 \pm 0.1$ \\
SCDM mock2 & 1 & 0.5 & 1.3 & 0 & 250
         &  $9.6 \pm 1.1$ & $1.2 \pm 0.1$ & $1.3 \pm 0.8$ & $1.2 \pm 0.1$ 
         & $14.6 \pm 1.1$ & $1.4 \pm 1.1$ \\
SCDMc mock3 & 1    & 0.5 & 0.6 & 0.2 & 900
         &  $5.1 \pm 1.2$ & $1.5 \pm 0.1$ & $1.0 \pm 1.5$ & $1.1 \pm 0.1$ 
         & $11.9 \pm 1.2$ & $1.4 \pm 0.1$ \\
$\Lambda$CDM mock4 & 0.3 & 0.65 & 1.2 & -0.9 & 600               
         &  $4.1 \pm 0.9$ & $1.9 \pm 0.1$ & $1.2 \pm 0.9$ & $1.2 \pm 0.1$ 
         &  $8.6 \pm 1.5$ & $2.2 \pm 0.1$ \\
$\Lambda$CDM mock5 & 0.3 & 0.65 & 1.2 & -0.5 & 300               
         &  $6.3 \pm 1.5$ & $1.8 \pm 0.1$ & $1.6 \pm 1.7$ & $1.2 \pm 0.2$ 
         & $11.9 \pm 1.7$ & $2.0 \pm 0.1$ \\
$\Lambda$CDM mock6 & 0.3 & 0.65 & 1.2 & 0    & 200               
         &  $5.8 \pm 1.5$ & $1.9 \pm 0.1$ & $2.7 \pm 0.6$ & $1.1 \pm 0.1$ 
         & $17.8 \pm 1.0$ & $1.5 \pm 0.1$ \\
$\Lambda$CDM mock7 & 0.3 & 0.65 & 1.2 & 1    & 100               
         &  $7.6 \pm 0.9$ & $1.5 \pm 0.1$ & $2.9 \pm 1.2$ & $1.1 \pm 0.1$ 
         & $14.2 \pm 1.2$ & $1.8 \pm 0.1$ \\
OCDM mock8 & 0.5 & 0.6 & 0.9 & 0 & 4000                       
         &  $6.6 \pm 0.5$ & $1.5 \pm 0.1$ & $2.5 \pm 0.4$ & $1.1 \pm .03$ 
         & $10.4 \pm 0.9$ & $2.0 \pm 0.1$ \\
OCDM mock9 & 0.5 & 0.6 & 0.9 & 1 & 500                        
         &  $5.2 \pm 1.4$ & $1.7 \pm 0.1$ & $4.5 \pm 0.9$ & $1.0 \pm 0.1$ 
         &  $5.9 \pm 0.9$ & $2.4 \pm 0.2$ \\
OCDM mock10& 0.5 & 0.6 & 0.9 & 2 & 300                        
         & $10.7 \pm 1.8$ & $1.2 \pm 0.2$ & $4.8 \pm 0.6$ & $1.0 \pm 0.1$ 
         & $15.7 \pm 2.7$ & $1.8 \pm 0.2$ \\
BCDM mock11& 1   & 0.5 & 0.6 & 0.5 & 20000                       
         &  $7.7 \pm 0.9$ & $1.5 \pm 0.1$ & $1.8 \pm 1.0$ & $1.2 \pm 0.1$ 
         & $10.2 \pm 3.6$ & $2.3 \pm 0.4$ \\
                              
\end{tabular}

\end{minipage}
\end{table*}

\begin{figure}
\label{fig4}
\psfig{figure=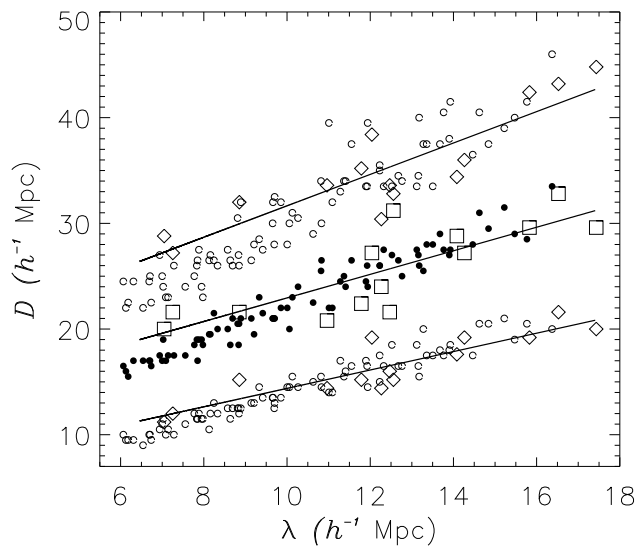,width=9cm}
\caption{\label{datvds} Median (squares) and quartile (diamonds) values of
the void sizes in the LCRS versus the mean galaxy separation $\lambda$ in 
the 14 volume limited data sets.  The straight lines give the linear fits 
of Eq. (2) to the data.  The small circles (filled for median and open for 
quartiles) show the corresponding void sizes in the mock sample 8 of the 
OCDM simulation (which has typically too small void sizes for small 
$\lambda$).}
\end{figure}

\begin{figure}
\label{fig5}
\psfig{figure=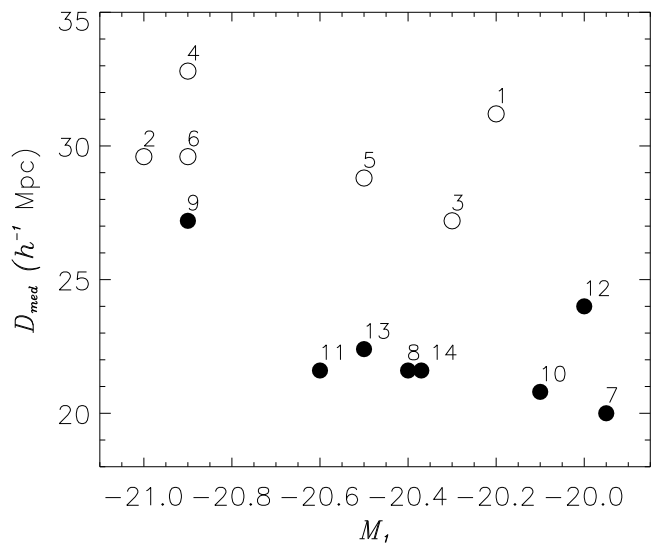,width=9cm}
\caption{\label{datvdsa} Median values of the void sizes in the LCRS 
versus the lower absolute magnitude limit $M_1$ in the 14 volume
limited data sets.  The filled symbols show data sets with a sampling rate
$f_{min} > 0.35$, the open symbols less well sampled data with $f_{min} <
0.35$.  The numbers at the symbols show the different data sets from Table
1.}
\end{figure}

\section{Voids in CDM mock samples}

For evaluation of our results we compare them with a set of numerical
simulations and corresponding mock samples of model galaxy
distributions in different CDM models.  We compare the model galaxy
distribution also with the 2-point correlation function of galaxies
with a similar range in brightness as that in the LCRS.

\subsection{Construction of the mock samples} 

We employ particle mesh (PM) simulations in different cosmological
models.  First, we consider a COBE normalized SCDM model with
$\Omega_m=1$ and dimensionless Hubble constant $h=0.5$.  For the COBE
normalization, we take the prescription of \citet{BW97}.  As an
alternative, we take the same model at an earlier time, SCDMc, which
fits the requirements of cluster normalization (see, e.g.,
\citet{ECF96}).  Further, we study more realistic models: $\Lambda$CDM
with $\Omega_m=0.3$, $h=0.65$, and a cosmological term to provide
spatial flatness; and an open model, OCDM with $\Omega_m=0.5$,
$h=0.6$.  Finally, we consider a high density CDM model with a more
complex initial spectrum -- one with a break in power between large
and small scales -- denoted as a broken scale invariant or BCDM model.
Earlier discussions of this model and an analytic fit to the spectrum
are given in \citet{KMGMR95}.  We perform simulations with $300^3$
particles in $600^3$ cells, and we simulate large boxes of $(500
\mhmpc)^3$ volume.  These simulations are described in more detail in
\citet{RBGKM98}, where we studied the cluster power spectrum, and in
\citet{MDRT98} and \citet{DMRT99}, where we simulated overdensity 
regions in the LCRS that correspond to superclusters of galaxies. 
Therefore, the present investigation should provide complementary 
information.  In using a large box size, we have a sufficient volume 
to simulate reliably voids with sizes of up to 60 \hmpc\/ and to find 
a reasonable representation of the void hierarchy. The price to be paid 
for these box sizes is a particle mass of $(1 - 3) \times 10^{11}$ \msun.  
In other words, we must identify galaxies with single mass points.

For galaxy identification, we employ the ideas of \citet{EJS80} to
differentiate the simulation particles in voids for low environmental
densities and in clustered galaxies for densities higher than a
critical density $\delta_{th}$.  To this aim, we determine the density
around each simulation particle at a fixed radius of 1 \hmpc.  Then,
we identify no galaxies if the local overdensity $\delta$ is smaller
than a threshold, $\delta < \delta_{th}$, and above, $\delta >
\delta_{th}$, we identify galaxies with a probability 
\be
P(\delta)=1-(\delta/\delta_{cr})^{1/3}. 
\ee 
A threshold was also used in \citet{Ei99} where the bias factor was
determined from the amount of matter in voids.  Furthermore, hydrodynamic
simulations of \citet{WHK97} and \citet{CO99} have confirmed that galaxy
formation is inefficient in low density regions where the density is
determined on galactic scale.  The probability distribution for $\delta >
\delta_{th}$ is only a slight modification which hinders a strong
increase of the galaxy clustering at high local densities.  Physically,
it models the merging of small galaxies in high density regions.  This
modification becomes mainly effective for models which are highly evolved
at galaxy scales, such as SCDM and $\Lambda$CDM.  For a discussion of
such a local bias prescription, compare also \citet{MPH98}.  A
two-parameter model for galaxy identification is similar to the method
used by \citet{CHWF98} to produce mock samples of the 2dF- and Sloan
Digital Sky surveys.  It was also employed for LCRS mock samples in
\citet{DMRT99}.  According to the motivation, we expect that the
threshold overdensity $\delta_{th}$ determines the size distribution of
voids in the mock samples.  We vary its value between $\delta_{th} = -0.9
\dots 2$, but with a prevalence of values near zero.  We also checked the
2-point correlation function of mock samples in comparison with the
redshift space correlation function of the LCRS galaxies given by
\citet{Tu97} and with a reconstruction of the real space correlation
function of similarly bright APM galaxies given by \citet{Ba96}.  For
reproducing the correlation function, the suppression of galaxy numbers
in high density regions as modeled by Eq.  (3) is important (see also
\citet{JMB98}).  The parameters of the 11 mock samples we use are given
in Table \ref{fit}.

In Fig. 6, we compare the correlation functions of one mock sample for
each CDM model with the data of APM galaxies in real space according
to \citet{Ba96}.  The correlation function of the SCDM mock samples 2
and 3 reproduce the correlation function in the highly clustered
region $r < r_0 \approx 5.5 \mhmpc$.  Here, $r_0$ denotes the
correlation length, and, at smaller scales, the correlation function
is well described by a power law $\xi = (r/r_0)^{-1.6}$.  The SCDM
models, however, cannot reproduce the correlation function at larger
radii.  It is well known that the SCDM model has insufficient power on
large scales to reproduce the observed clustering of galaxies.  There
is no possibility to cure this difficulty with our simple bias
prescription.  The correlation function of mock sample 1 is not shown,
but it looks similar to that of mock sample 2.

The correlation function of mock sample 7 for the $\Lambda$CDM and of
the mock sample 8 for the OCDM model reproduce the correlation
function between $1 \mhmpc < r_0 < 40 \mhmpc$.  At small separations,
they stay below the observed values.  This is due to the poor spatial
resolution of our PM simulations, but it has no influence on the void
statistics at the large scales studied in this paper.  Actually, the
mock sample 4 for the $\Lambda$CDM model delivers a slightly better
correlation function than the sample shown (sample 7).  The mock
samples 5, 6, and 7 for the $\Lambda$CDM model and the two mock
samples 9 and 10 for the OCDM are produced to test whether a higher
threshold $\delta_{th}$ can improve the void statistic of these
models.  In fact, the correlation function of mock samples 5 to 7 for
$\Lambda$CDM are almost as good as that of mock sample 4, whereas the
mock samples 9 and 10 of the OCDM model lead to correlation functions
lying below the observed one at large scales, $r > 10 \mhmpc$.
Obviously, the high threshold in this model leads to a strong
suppression of the mock galaxy density in medium density regions, and
therefore to a suppression of the correlation function on these scales. 
As Fig. 6 demonstrates, the mock sample 11 of the BCDM model leads to 
a good fit of the correlation function over the total range shown. The 
LCRS correlation function in redshift space of \citet{Tu97} is 
similarly fitted (see a similar comparison with CDM models in that 
paper).

\begin{figure}
\label{fig6}
\psfig{figure=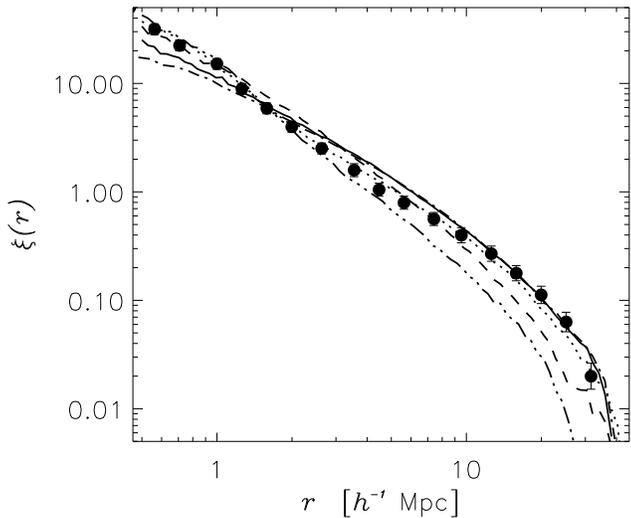,width=9cm}
\caption{Two-point correlation function of mock 2 for SCDM (dashed),
of mock 3 for SCDMc (dash-dot-dot-dotted), of mock 7 for $\Lambda$CDM
(solid), of mock 8 for OCDM (dash-dotted), and of mock 11 for BCDM
(dotted) as compared with the data of \citet{Ba96}.}
\end{figure}

\subsection{Voids in mock CDM models}

Voids in the mock samples are found with the same algorithm as in the
LCRS data.  Here we search for voids both in square areas of the
simulation box and in volumes representing a similar wedge-like
geometry as in the LCRS data.  For the mock prescription 6 in the OCDM
model, we also selected 10 realizations with a wedge geometry.  To
this aim, we place a fictitious observer in one corner of the
simulation box and produce a wedge-like section of $80 \times 1.5$
degree extension, as in the LCRS slices, and select the galaxies in
the distance range between 200 and 350 \hmpc.  We always take the
galaxies in redshift space for comparison with the data, but this is
little difference between the void statistics of the real space and of
the redshift space results.  Finally, we select randomly reduced
subsets of the 11 mock galaxy catalogues to study the dependence of
the void statistics on the galaxy density.

The restriction of the mock samples to the survey geometry leads to
cumulative void distributions that are slightly shifted (less than
10\%) to larger void sizes $D$ (recall the discussion in Section 2.2).
More important is the cosmic variance which is evident in the medians
and the quartiles in Fig. 4.  It amounts to about 2 \hmpc\/ for the
median and about 3 \hmpc\/ for the upper quartile values.  The
increase of the variance at larger void sizes is a trivial consequence
of the smaller number of large voids.  It becomes already obvious from
the cumulative void size distribution of the LCRS data in Fig. 3.

The large number of independent realizations of the void statistics
from the OCDM mock sample 8 shown in Fig. 4 makes it obvious that the
lower quartile of the mock samples can reproduce the LCRS data, but
the median and particularly the upper quartile are systematically too
low.  The fits of the relation Eq. (2) to the models are shown in
Table \ref{fit}.  The median and 75th percentile of the residual void
sizes $D_0$ of the OCDM mock sample 6 are about $2 \sigma$ beyond the
well sampled LCRS data.  The mock samples 9 and especially 10 can
better reproduce the data, but the bias prescription in these models
leads to an insufficient fit of the correlation function.  The results
of the fits from all studied models are collected in Table \ref{fit}.
The $\Lambda$CDM mock samples can reproduce the void size distribution
if a bias threshold $\delta \approx 1$ is imposed.  The SCDM models
are even worse than the OCDM models.  A reasonable representation of
the void data is yielded by the BCDM mock sample 11, which also provides 
a good representation of the correlation function.

The scaling relations as illustrated in Fig.  4 along with numerical
values given in Table \ref{fit} are the main result of our study.
Obviously a realistic cosmological model, such as the $\Lambda$CDM or the
BCDM model, and a suitable phenomenological bias prescription can
reproduce the void distribution in the LCRS slices.  It should be noted
that the quality of the fits of relation Eq.  (2) to the data has some
uncertainties.  This becomes also obvious from a visual inspection of
Fig.  4.  Therefore, similar deep data sets as the LCRS in a less
restrictive geometry are very promising in sharpening the cosmological
conclusions from this analysis of the void statistics.

Beyond the study of the quartiles, the complete void distribution
contains information on the degree of variance in the data and in the
mock samples.  In Fig.  7, we show the cumulative void distributions of
two data sets, 7 and 14, which are characterized by a similar mean galaxy
separation of $7 \lta \lambda \lta 7.5$ \hmpc.  Similarly, the cumulative
void size distribution of mock samples with a similar mean galaxy
separation for the OCDM model is shown as solid line with $1 \sigma$
error bars.  It shows about 3 \hmpc\/ smaller voids over the complete
distribution function, but the significance is only about $2 \sigma$.
The $\Lambda$CDM model (mock 7) and the BCDM model (mock 11) look much
better.  The small discrepancy concerns only the few largest voids in the
data.  Not unexpectedly, the void distribution of the Poisson samples
strongly underestimates the cumulative void distribution.

It is remarkable that the void size distribution of data set 14, which
is inhomogeneously sampled, lies almost on top of that of the
homogeneously sampled set 7.  Obviously, the different sampling rates
reduce the galaxy distribution mostly in the highly clustered areas,
and there is less influence on the medium density regions studied by
the void size distribution.

\begin{figure}
\label{fig7}
\psfig{figure=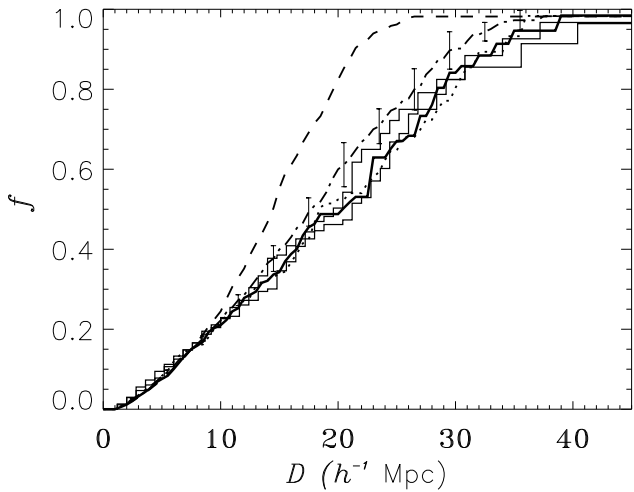,width=9cm}
\caption{Cumulative fraction of area $f$ covered by voids as function of
the size for the data sets 7 and 14, compared with mock 7 ($\Lambda$CDM) as solid
line, mock 8 (OCDM) as dash-dotted line with $1 \sigma$ error bars, mock 11
(BCDM) as dotted line, and of a Poisson sample with the same density as
dashed line.}
\end{figure}

\section{Discussion}

The present paper represents the first analysis of the void
distribution in the LCRS.  The LCRS gives a unique prospect for
studying the statistics of voids in the galaxy distribution.  In
distinction from the correlation function, thereby we are sensitive to
the galaxy distribution in medium density regions which are visually
characterized by the occurrence of a hierarchy of filaments and small
pancakes.  We employed strict selection criteria to get 13
homogeneously diluted volume limited data sets selected with different
magnitude limits in different parts of the survey.  As Fig.  3
demonstrates, the different cumulative distributions of the area
coverage of voids look very similar, independent of the sample size.
Therefore, we can speak of a hierarchy of voids that characterizes the
part of space which is devoid of galaxies.  This hierarchy ends at
about 40\hmpc, a typical value of the maximum void size in well
sampled parts of the LCRS as seen from Table \ref{data}.  It is much
smaller than the total size of the observed samples.

The similarity in the different void size distributions becomes also
obvious if we employ the ratio of the void size to the median void
size of the sample $\mu = D/D_{med}$ as an independent variable
(Fig. 8).  The well sampled data sets 7 and 8 deliver almost
coinciding distributions.  The remaining ones lead to a larger
scatter, especially in the part of the survey covered by voids larger
than the median.  The solid curve provides a simple fit of this mean
behavior of the cumulative void distribution.  We also show the
differential void distribution as a solid line with a broad maximum
near the median void size $D_{med}$.  For comparison, the dashed line
shows a similar fit to the void distribution of random points which is
much more peaked at the median void size.  Typically, the comparison
of the data with the Poisson samples show both a higher probability of
voids larger and of voids smaller than the median.  This bimodality in
the observed void distribution is a result of the gravitational
instability on a wide range of scales that is typical for CDM models.
Qualitatively, it is well reproduced by all our CDM simulations.
Large voids are typical for regions empty of the highly dense clusters
and superclusters of galaxies.  The high probability of small voids
shows that the gravitational clustering with its typical appearance of
a hierarchy of filaments and sheets extends to small scales, and that
small voids are very abundant in this hierarchy.  They have dimensions
similar to the mean galaxy separation in our observed samples, and
they enter the medium density regions of the supercluster
distribution.

A quantitative comparison of the void size distribution of the data with
CDM models, as shown in Table \ref{fit}, demonstrates that the size
distribution of the small voids is well reproduced by all our different
mock samples.  Much more critical is the quantitative comparison of the
void size distribution of the voids larger than the median size.  Despite
the remarkable scatter, it becomes obvious that most mock samples have
difficulty in reproducing the abundance of large voids.  Sufficient void
distributions are obtained in the $\Lambda$CDM model for a bias threshold
$\delta_{th} = 1$, the mock sample 10 of the OCDM model with a high
threshold $\delta_{th} = 2$, and the mock sample 11 of the BCDM model.
The latter must be strongly biased due to the reduced power at galaxy
scales in this model.  The OCDM mock sample 10 has the difficulty of
insufficient amplitude of the two-point autocorrelation function at large
scales.

The main aim of the present study was to show that the void statistics of
the LCRS provides an interesting and sensitive cosmological test of the
galaxy formation.  It is clear from our analysis that it probes larger
scales than the two-point correlation function, and that it is a reliable
statistic with additional information on the nature of the galaxy
distribution in a low density environment.

\begin{figure}
\label{fig8}
\psfig{figure=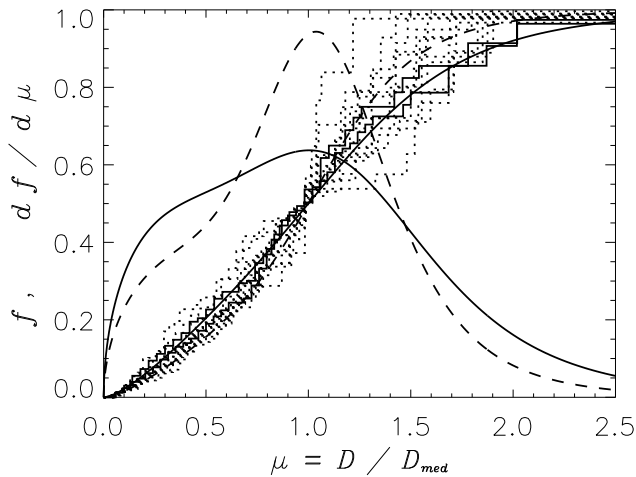,width=9cm}
\caption{Cumulative and differential distributions, $f$ and $d f
/d \mu$, of the LCRS areas covered by voids as function of the
normalized void size $\mu = D/D_{med}$.  The solid histograms show the
better sampled data sets 7 and 8, whereas the dotted ones stem from
the diluted galaxy samples that scatter especially for large values of
$\mu$.  The monotonicly increasing solid curve provides a fit of the
mean behavior, the monotonic dashed lines shows the behavior of
Poisson samples.  The peaked solid and dashed lines are differential
size distributions from the data and from a set of Poisson samples,
respectively.}
\end{figure}

A similarly important result of the present study is the derivation of
a simple scaling relation (2) of the void sizes that give a cumulative
voids coverage of 25\%, 50\%, and 75\% in the galaxy distribution as
shown in Fig. 4.  The residual void size $D_0 \approx 12 \pm 3$
\hmpc\/ for the mean and the increase rate $\nu \approx 1$ in diluted
samples characterize the hierarchy of voids in the LCRS.  The void
size distribution depends much more on the mean galaxy density than on
the size of the survey volume or on the absolute magnitude of the
galaxy sample.  In fact, we only discussed the density dependence
since only this dependence can be derived from the given data sets
with high statistical significance.  The strong suppression of the
voids with sizes $D > 2.5 D_{med}$ underlines the fact that the size
of the LCRS is large enough to get a reasonable estimate of the
abundance of large voids and indicates a transition of the galaxy
distribution to a homogeneous distribution at larger scales.

A further important point concerns the independence of the void size
distribution with regard to the absolute magnitude range of the
selected galaxies -- e.g., the comparison of the void sizes in the
data sets 1, 9, 11, 12, and 13.  There, the median void sizes are
similar, and only the maximum void sizes differ by about 15\%.  We
ascribed it to the different depth of the data sets, but we should
keep in mind that a larger volume corresponds in general to brighter
galaxies.  Disentangling both effects requires the extension of
redshift surveys to a larger magnitude range, where even a small
extension promises some progress as the comparison of the fields
sampled partly with the 50 fiber spectrograph and the data from the
$-12 \dgr$ slice which is sampled completely with the 112 fiber
spectrograph (sets 7, 8, 9, and 14) demonstrates.

The self-similarity of the void distribution in the LCRS is
illustrated in Figs. 3 and 7, which show quite comparable shapes.  The
key dependence on the mean galaxy density is illustrated in Fig. 4 and
in the two-parameter fit of the median and quartile void size
distributions by the relations Eq. (2).  Such relations are well
reproduced by the hierarchical clustering in CDM models as the
parameters in Table \ref{fit} demonstrate.  The void distribution
provides a sensitive test of these models.  The dependence of the 
void statistics on the threshold criterion for the mock galaxy 
indentification shows that the galaxy biasing has a stronger influence 
than do cosmological parameters like the mean matter density or the 
precise form of the primordial power spectrum.

\bibliographystyle{mnras}
\bibliography{void4}

\end{document}